\documentclass[12pt,english]{article}
\usepackage[T1]{fontenc}
\usepackage[latin1]{inputenc}
\usepackage{amssymb}

\makeatletter

\usepackage[english]{babel}
\DeclareInputText{"0AD}{\textendash}

\ProvideTextCommandDefault{\quotedblbase}{%
\raisebox{-1ex}{\textquotedblright}
\hspace{-0.7em}
}
\ProvideTextCommandDefault{\quotesinglbase}{%
\raisebox{-1ex}{\textquoteright}
\hspace{-0.7em}
}
\ProvideTextCommandDefault{\guillemotleft}{%
\raisebox{0.27ex}{\ensuremath{\scriptscriptstyle \ll\!\!\!}}
}
\ProvideTextCommandDefault{\guillemotright}{%
\raisebox{0.27ex}{\ensuremath{\scriptscriptstyle \gg}}
}
\ProvideTextCommandDefault{\guilsinglleft}{%
\raisebox{0.27ex}{\ensuremath{\scriptscriptstyle <\!\!\!}}
}
\ProvideTextCommandDefault{\guilsinglright}{%
\raisebox{0.27ex}{\ensuremath{\scriptscriptstyle >}}
}
\ProvideTextCommandDefault{\DH}{%
D\hspace{-0.7em}\rule[0.8ex]{0.30em}{0.08ex}\hspace{0.40em}
}
\ProvideTextCommandDefault{\dh}{%
\ensuremath{\mathrm{\partial}}
\hspace{-0.65em}\rule[1.35ex]{0.3em}{0.08ex}\hspace{0.35em}
}
\ProvideTextCommandDefault{\TH}{%
\textsc{I\hspace{-0.325em}p}
}
\ProvideTextCommandDefault{\th}{%
p\hspace{-0.55em}l
}

\usepackage{babel}
\makeatother
\begin{document}

\title{Spin gap antiferromagnets: materials and phenomena}

\author{Indrani Bose\\
Department of Physics\\
Bose Institute\\
93/1, A. P. C. Road\\
Kolkata-700009}

\maketitle
\begin{abstract}
There are several interacting spin systems which have a gap in their
spin excitation spectrum. The gap does not occur due to anisotropies
present in the system but is quantum mechanical in origin. We give
a brief overview on the different types of spin gap (SG) antiferromagnets,
the models proposed to describe their physical properties and experimental
realizations of such systems. Our special focus is on exactly-solvable
models and rigorous theories which provide the correct physical picture
for the novel phenomena exhibited by SG systems. 
\end{abstract}

\newpage

The last decade has witnessed an unprecedented research activity on
undoped and doped quantum antiferromagnets (AFMs) in low dimensions.
Several new materials exhibiting a variety of novel phenomena have
been discovered. The experimental effort is closely linked with theoretical
ideas. The field of low-dimensional quantum magnetism provides a fertile
ground for rigorous theory. Powerful techniques like the Bethe Ansatz
(BA)$^{1}$ and bosonization are available to study ground and excited
state properties. Models of interacting spin systems are known for
which the ground state and in some cases the low-lying excitation
spectrum are known exactly. Theorems, furthermore, offer important
insight on the nature of ground and excited states. The knowledge
gained provides the impetus to look for real materials so that experimental
confirmation of theoretical predictions can be made. In this review,
we discuss a special class of low-dimensional AFMs, the so-called
spin gap (SG) systems, to illustrate the rich interplay between theory
and experiments. The review is not meant to be exhaustive but focuses
on some broad classes of phenomena exhibited by SG systems.

The SG AFMs are characterised by a gap in the spin excitation spectrum.
The magnitude of the gap, $\Delta$, is measured as the difference
between the energies of the lowest excited state and the ground state.
The excitation spectrum of an AFM is said to be gapless if $\Delta=0.$
In general, a gap appears in the spin excitation spectrum if some
kind of anisotropy is present in the system. In SG AFMs, however,
the gap is purely quantum mechanical in origin and cannot be ascribed
to any anisotropy. The basic spin-spin interactions in AFMs are described
by the Heisenberg Hamiltonian

\begin{equation}
H=\sum_{<ij>}J_{ij}\mathbf{S}_{i}.\mathbf{S}_{j}\label{1}\end{equation}

\noindent where $\mathbf{S}_{i}$ is the spin operator located at
the lattice site $i$ and $J_{ij}$ denotes the strength of the exchange
interaction. Real magnetic materials are three-dimensional (3d) but
behave effectively as low-dimensional systems if the dominant exchange
interactions are intra-chain (1d) or intra-planar (2d). For most materials,
the exchange interaction is confined to only the nearest-neighbour
(n.n.) spins on the lattice. Also, $J_{ij}$'s have the same magnitude
$J$ for all the n.n. interactions. There are, however, spin systems
in which further-neighbour interactions cannot be ignored and the
exchange interaction strengths are inhomogeneous in character. In
the next section, we discuss some possible origins of SG, the models
proposed to describe such systems and experimental evidence for the
different mechanisms of SG.

\subsection*{\noindent {\Large Origins of Spin Gap}}

SG AFMs have spin-disordered ground states, i.e., the spin-spin correlations
in the ground state are short-ranged. The ground states, in the absence
of long range magnetic order, can broadly be described as quantum
spin liquids (QSLs). The spin liquids are distinct from simple paramagnets,
in certain cases one can define novel order parameters which have
non-zero expectation values in the QSL states. Formation of such states
is favoured by quantum fluctuations the effect of which is prominent
in low dimensions and for low values of the spin. In experiments,
the presence of the gap $\Delta$ is confirmed through measurement
of quantities like the susceptibility, $\chi,$ which goes to zero
exponentially at low T as $\chi\sim exp\left(-\frac{\Delta}{k_{B}T}\right)$.
Some well-known examples of SG AFMs are: spin-Peierls (SP) systems,
AFM compounds consisting of weakly-coupled spin dimers, frustrated
spin systems, spin ladders and Haldane gap (HG) AFMs. The SP compounds
are historically the first examples of magnetic systems exhibiting
a SG. The SP transition was originally observed in some organic compounds.
In 1993, Hase et al.$^{2}$ obtained the first experimental evidence
of the SP transition in an inorganic compound, $CuGeO_{3}.$ The SP
transition generally occurs in quasi-one dimensional (1d) AFM spin
systems with half-odd integer spins and is brought about by spin-phonon
coupling. Below the SP transition temperature $T_{SP}$, a periodic
deformation of the lattice sets in such that the distances between
neighbouring spins are no longer uniform but alternate in magnitude.
This results in an alternation, $J(1+\delta)$and $J(1-\delta),$
in the strengths of the n.n. exchange interaction strengths. The ground
state is dimerized in which singlet spin pairs occupy the links with
enhanced exchange couplings. The SP ground state is spin disordered
and a finite energy gap exists in the $S=1$ spin excitation spectrum.

Some AFM compounds can be described as crystalline networks of spin
dimers. Well-known examples of such systems are the compounds $ACuCl_{3}$
( $A=K,Tl$ ) in which a spin dimer arises from two antiferromagnetically
coupled $Cu^{2+}$ions$^{3,4}$. The dimer ground state is a spin
singlet with total spin $S=0.$ The excited triplet state with total
spin $S=1$ is separated from the ground state by an energy gap. The
excitation created on a particular dimer propagates through the network
of dimers due to the weak inter-dimer exchange coupling. 

\noindent \textbf{\large Frustrated spin systems}{\large \par}

Quantum fluctuations in a spin system are enhanced due to the presence
of frustration in the system. Frustration implies a conflict in minimising
the exchange interaction energies associated with different spin pairs.
Such conflicts arise mainly due to lattice topology and the presence
of competing further-neighbour interactions$^{5}$. AFM systems with
an odd number of bonds in the unit cell of the underlying lattice
are frustrated spin systems. Examples include the triangular, kagom\'{e},
pyrochlore and FCC lattices. Consider three Ising spins located at
the vertices of a triangular plaquette and interacting antiferromagnetically
with each other. The energy of an interacting spin pair is minimised
when the spins are antiparallel. The three interacting spin pairs
cannot, however, be simultaneously made antiparallel so that one pair
is in the parallel spin configuraton in the lowest energy state. Such
conflicts are absent when the elementary plaquette of the lattice
contains an even number of bonds as in the case of the square lattice.
Determination of the ground state of the AFM Heisenberg exchange interaction
Hamiltonian (eq.(1)) on various lattices poses a formidable theoretical
problem. The task becomes simpler if the spins are treated as classical
vectors, i.e., the magnitude of spins $S\rightarrow\infty.$ The energy
of a triangle of spins is given by 

\begin{equation}
E_{t}=\frac{J}{2}\left[(\mathbf{S}_{1}+\mathbf{S}_{2}+\mathbf{S}_{3})^{2}-(S_{1}^{2}+S_{2}^{2}+S_{3}^{2})\right]\label{2}\end{equation}

\noindent The ground state, i.e., the minimal energy spin configurations
are obtained for $\mathbf{S}_{tot}=\mathbf{S}_{1}+\mathbf{S}_{2}+\mathbf{S}_{3}=0.$
In the case of the full triangular lattice, the classical ground state
has a three-sublattice order with each elementary plaquette of spins
satisfying the constraint $\mathbf{S}_{tot}=0.$ In the case of the
kagom\'{e} lattice, which is constructed out of triangular plaquettes,
the classical ground state is infinitely degenerate$^{6}$. In the
first case, quantum fluctuations do not fully destroy the order in
the ground state but reduce the sublattice magnetization significantly
from its classical value. Recent calculations show that the $S=\frac{1}{2}$
Heisenberg AFM (HAFM) Hamiltonian on the triangular lattice has long
range order (LRO) in the ground state$^{7,8}$. A rigorous proof of
the existence of LRO is, however, still missing. Experimental realizations
of the triangular lattice HAFM include $VCl_{2},VBr_{2},C_{6}Eu,NaNiO_{2}$
etc$^{9}$. In the case when the classical ground state is highly
degenerate, i.e., spin-disordered, thermal/quantum fluctuations may
select a subset of states as ground states leading to new forms of
spin order. This is the phenomenon of `order from disorder'$^{10}$.
Another possibility is the opening up of a gap in the spin excitation
spectrum. The $S=\frac{1}{2}$ HAFM on the kagom\'{e} lattice appears
to be a SG system with a singlet-triplet gap in the spin excitation
spectrum. An interesting feature of the SG system is the existence
of a large number of singlet excitations in the gap, the number of
which is proportional to $(1.15)^{N},$where N is the number of sites
in the lattice$^{11,12}$. The temperature dependence of the susceptibility
is not affected by the presence of singlet excitations but the specific
heat possibly has a power-law dependence, $C_{v}\varpropto T^{\alpha},$
due to contributions from the singlets. An experimental realization
of a $S=\frac{1}{2}$ HAFM on the kagom\'{e} lattice is yet to be
obtained. The compound $SrCr_{9}Ga_{12}O_{19}$$^{5}$ is an example
of a kagom\'{e}-lattice AFM with $S=\frac{3}{2}.$ The dominant contribution
to the low temperature specific heat of this compound appears to come
from the singlet states. 

The $S=\frac{1}{2}$ $J_{1}-J_{2}$ model in 1d, describing the Majumdar-Ghosh
(MG) chain$^{13},$ is the first example of a frustrated quantum spin
model with further-neighbour interactions for which the ground state
can be determined exactly. The Hamiltonian with periodic boundary
conditions (PBC) is given by

\begin{equation}
H_{MG}=J_{1}\sum_{i=1}^{N}\mathbf{S}_{i}.\mathbf{S}_{i+1}+J_{2}\sum_{i=1}^{N}\mathbf{S}_{i}.\mathbf{S}_{i+2}\label{3}\end{equation}
\noindent where $J_{1}$ and $J_{2}$ are the n.n. and next-nearest-neighbour
(n.n.n.) exchange interaction strengths. The exactly-solvable MG point
corresponds to $\frac{J_{2}}{J_{1}}=\frac{1}{2}.$ The exact ground
state is doubly degenerate and the ground states are

\begin{equation}
\phi_{1}\equiv\left[12\right]\left[34\right]\left[56\right]......\left[N-1N\right],\phi_{2}\equiv\left[23\right]\left[45\right]\left[67\right]......\left[N1\right]\label{4}\end{equation}

\noindent where $\left[lm\right]$ denotes a singlet spin configuration,
$\frac{1}{\sqrt{2}}\left(\alpha(l)\beta(m)-\beta(l)\alpha(m)\right),$
for spins located at the lattice sites $l$ and $m$. The up and down
spin states are denoted as $\alpha$ and $\beta$. A singlet state
is also known as a valence bond (VB). The excitations in the model
can be described in terms of scattering spin-$\frac{1}{2}$ defects
acting as domain walls between the two exact ground states$^{14}.$
The scattering states form a continuum, the lowest branch of which
is separated from the ground state by a gap. The MG chain is thus
a SG AFM. The SG phase survives for $\alpha=\frac{J_{2}}{J_{1}}$
greater than a critical value $\alpha_{cr}\simeq0.2411.$ For $0<\alpha<\alpha_{cr},$
a gapless phase is obtained. If the n.n. exchange interactions are
alternating in strength, as in a SP system, the ground state is nondegenerate
with the VBs forming along the stronger bonds. The MG Hamiltonian
has been used to study the properties of the SP compound $CuGeO_{3}.$
A large number of studies has been carried out on the frustrated $J_{1}-J_{2}$
and $J_{1}-J_{2}-J_{3}$ models in 2d where $J_{2}$ denotes the strength
of the diagonal exchange couplings and $J_{3}$ that of n.n.n. interactions
in the horizontal and vertical directions. The most general model
studied so far is the $J_{1}-J_{2}-J_{3}-J_{4}-J_{5}$ model$^{15}.$
$J_{4}$ and $J_{5}$ are the strengths of the knight's-move-distance-away
and further-neighbour diagonal exchange interactions respectively
(Figure 1). The four columnar dimer (CD) states (Figure 2) are the
exact eigenstates of the $J_{1}-J_{2}-J_{3}-J_{4}-J_{5}$ Hamiltonian
for the ratio of interaction strengths 

\begin{equation}
J_{1}:J_{2}:J_{3}:J_{4}:J_{5}=1:1:\frac{1}{2}:\frac{1}{2}:\frac{1}{4}\label{5}\end{equation}
\noindent Each dotted line in Figure 2 represents a VB, i.e., a singlet
spin configuration. There is no rigorous proof as yet that the CD
states are the exact ground states though approximate theories tend
to support the conjecture$^{16}.$ One can, however, prove that any
one of the CD states is the exact ground state when the dimer bonds
are of strength $7J$ and the rest of the exchange interactions are
of strengths as specified in eq. (5). The excitation spectrum of the
model is separated from the ground state by a gap.

The Shastry-Sutherland (SS) model$^{17}$ is an example of a frustrated
SG AFM in 2d. Figure 3 shows the lattice on which the model is defined.
The n.n. and diagonal exchange interactions are of strengths $J_{1}$
and $J_{2}$ respectively. For $\frac{J_{1}}{J_{2}}$ less than a
critical value $\simeq$ 0.7, the exact ground state consists of singlets
along the diagonals. At the critical point, the ground state changes
from the gapped disordered state to an antiferromagnetically ordered
gapless state. The AFM compound $SrCu_{2}(BO_{3})_{2}$ is an experimental
realization of the SS model$^{18}.$ Triplet excitations in the model
are found to be almost localized. A single triplet excitation can
propagate in the SS lattice only at the sixth order perturbation in
$\frac{J_{1}}{J_{2}}.$ 

\noindent \textbf{\large Spin ladders}{\large \par}

Spin ladders constitute one of the most well-known examples of SG
AGMs. The simplest ladder model consists of two chains coupled by
rungs (Figure 4). In general, the ladder may consist of $n$ chains.
In the spin ladder model, each site of the ladder is occupied by a
spin (usually of magnitude $\frac{1}{2}$) and the spins interact
via the HAFM exchange interaction Hamiltonian (eq. (1)). The n.n.
intra-chain and rung exchange interactions are of strengths $J$ and
$J_{R}$ respectively. When $J_{R}=0,$ one obtains two decoupled
AFM spin chains for which the excitation spectrum is known to be gapless.
For all $\frac{J_{R}}{J}>0,$ a gap opens up in the spin excitation
spectrum$^{19}.$ The result is easy to understand in the simple limit
in which the exchange coupling $J_{R}$ along the rungs is much stronger
than the coupling along the chains. The intra-chain coupling in this
case may be treated as a perturbation. When $J=0,$ the exact ground
state consists of singlets along the rungs. The ground state energy
is $-\frac{3J_{R}N}{4}$, where $N$ is the number of rungs in the
ladder. The ground state has total spin $S=0.$ In first order perturbation
theory, the correction to the ground state energy is zero. An $S=1$
excitation may be created in the ladder by promoting one of the rung
singlets to the $S=1$ triplet state. The weak coupling along the
chains gives rise to a propagating $S=1$ magnon. In first order perturbation
theory, the dispersion relation is

\begin{equation}
\omega(k)=J_{R}+Jcosk\label{6}\end{equation}
\noindent where $k$ is the momentum wave vector. The SG is given
by

\begin{equation}
\Delta=\omega(\pi)\simeq J_{R}-J\label{7}\end{equation}
\noindent The two-spin correlations decay exponentially along the
chains showing that the ground state is a QSL. The family of compounds
$Sr_{n-1}Cu_{n+1}O_{2n}$ consists of planes of weakly-coupled ladders
of $\frac{(n+1)}{2}$ \noindent chains$^{20}.$ For $n=3$ and $5,$
one gets the two-chain and three-chain ladder compounds $SrCu_{2}O_{3}$
and $Sr_{2}Cu_{3}O_{5}$ respectively. The first compound is a SG
AFM while the second compound has properties similar to those of the
HAFM Hamiltonian in $1d$, which has a gapless excitation spectrum.
The experimental evidence is consistent with the theoretical prediction
that in an $n$- chain ladder, the excitation spectrum is gapped (gapless)
when $n$ is even (odd)$^{19}.$ Bose and Gayen$^{21}$ have studied
a two-chain ladder model with frustrating diagonal couplings. The
intra-chain and diagonal couplings are of equal strength $J$. For
$J_{R}\geq2J,$ the exact ground state consists of singlets along
the rungs with the energy $E_{g}=-\frac{3J_{R}N}{4}.$ An excitation
can be created by replacing one of the singlets by a triplet. The
triplet excitation is localised and separated by an energy gap from
the ground state. Xian$^{22}$ later pointed out that as long as $\frac{J_{R}}{J}>\left(\frac{J_{R}}{J}\right)_{c}\simeq1.401,$
the rung dimer state is the exact ground state.

Ladders provide a bridge between $1d$ and $2d$ many body systems
and are ideally suited to study how the electronic and magnetic properties
change as one goes from a single chain to the square lattice limit.
The significant interest in $2d$ many body systems is due to the
unconventional properties of the $CuO_{2}$ panes in doped cuprate
systems. The latter exhibit high temperature superconductivity in
appropriate ranges of dopant concentration. Many of the unusual properties
of the cuprate systems arise due to strong correlation effects. Ladders
are simpler systems in which some of the issues related to strong
correlation can be addressed in a rigorous manner. Doped ladder models
are toy models of strongly correlated systems. In these systems, strong
Coulomb correlations prohibit the double occupancy of a site by two
electrons, one with spin up and the other with spin down. In a doped
spin system, two processes are in competition: hole delocalization
and exchange energy minimization. The latter is minimised in the antiferromagnetically
ordered N\'{e}el-type of state. A hole moving in such a background
gives rise to parallel spin pairs which raise the exchange interaction
energy of the system. The questions of interest are: whether a coherent
motion of the holes is possible, whether two holes can form a bound
state and superconducting (SC) pairing correlations develop etc. These
issues are of significant relevance in the context of doped cuprate
systems in which charge transport occurs through the motion of holes.
In the SC phase, the holes form bound pairs with possibly $d-$wave
symmetry. Several proposals have been made so far on the origins of
hole binding but the actual binding mechanism is still controversial$^{23}.$
The doped cuprate systems exist in a `pseudogap' phase before entering
the SC phase. In fact, some of the cuprate systems also exhibit a
SG. The two-chain AFM ladder systems are SG systems and it is of interest
to study how the gap evolves on doping. The possibility of binding
of hole pairs in a two-chain ladder system was first pointed out by
Dagotto et al.$^{24}.$ In this case, the binding mechanism can be
understood in a simple physical picture. Again, consider the case
$J_{R}\gg J,$ i.e., a ladder with dominant exchange interactions
along the rungs. In the ground state, the rungs are mostly in singlet
spin configurations. On the introduction of a single hole, a singlet
spin pair is broken and the corresponding exchange interaction energy
is lost. When two holes are present, they prefer to be on the same
rung to minimise the loss in the exchange interaction energy. The
holes thus form a bound pair. In the more general case, detailed energy
considerations show that the two holes tend to be close to each other
effectively forming a bound pair. For more than two holes, several
calculations suggest that considerable SC pairing correlations develop
in the system. The superconducting state can be achieved only in the
bulk limit. Theoretical predictions motivated the search for ladder
compounds which can be doped wth holes. Much excitement was created
in 1996 when the doped ladder compound $Sr_{14-x}Ca_{x}Cu_{24}O_{41}$
was found to become SC under pressure at $x=13.6^{25}.$ The transition
temperature $T_{c}$ is $\sim12K$ at a pressure of $3GPa.$ As in
the case of cuprate systems, bound pairs of holes are responsible
for charge transport in the SC phase. Experimental results on doped
ladder compounds point out strong analogies between the doped ladder
and cuprate systems$^{26}.$ Bose and Gayen$^{21,27,28}$ have derived
exact, analytical results for the ground state energy and the low-lying
excitation spectrum of the frustrated $t-J$ ladder model doped with
one and two holes. The undoped frustrated ladder model has already
been described. The $t-J$ Hamiltonian describing the ladder is given
by

\begin{equation}
H_{t-J}=-\sum_{\left\langle i,j\right\rangle ,\sigma}t_{ij}\left(\widetilde{c}_{i\sigma}^{+}\widetilde{c}_{j\sigma}+H.C.\right)+\sum_{ij}J_{ij}\mathbf{S}_{i}.\mathbf{S}_{j}\label{8}\end{equation}
\noindent The $\widetilde{c}_{i\sigma}^{+}$ and $\widetilde{c}_{i\sigma}$
are the electron creation and annihilation operators which act in
the reduced Hilbert space (no double occupancy of sites),

\begin{equation}
\widetilde{c}_{i\sigma}^{+}=c_{i\sigma}^{+}(1-n_{i-\sigma}),\widetilde{c}_{i\sigma}=c_{i\sigma}(1-n_{i-\sigma})\label{9}\end{equation}
\noindent where $\sigma$ is the spin index and $n_{i},n_{j}$ are
the occupation numbers of the $i$th and $j$th sites. The first term
in eq. (8) describes the motion of holes. The hopping integral $t_{ij}$
has the value $t_{R}$ for hole motion along the rungs and the value
$t$ for both the intra-chain n.n. and diagonal hops. The latter assumption
is crucial for the exact solvability of the eigenvalue problem in
the one and two hole sectors. Though the model differs from the standard
$t-J$ ladder model$^{29,30}$ ( diagonal couplings missing in the
latter model ), the spin and charge excitation spectra exhibit similar
features. For the frustrated $t-J$ ladder, the dispersion relation
of the two-hole bound state branch can be obtained exactly and analytically.
The exact two-hole ground state is a bound state with centre of mass
momentum wave vector $K=0$ and $d$-wave type symmetry. The ladder
exists in the Luther-Emery phase in which the spin excitation is gapped
and the charge excitation gapless. There is no spin-charge separation,
a feature associated with Luttinger Liquids in which both the spin
and charge excitations are gapless. In the exact hole eigenstates,
the hole is always accompanied by a spin-$\frac{1}{2}.$ The hole-hole
correlation function can also be calculated exactly. When $J_{R}\gg J,$
the holes of a bound pair are predominantly on the same rung. For
lower values of $J_{R},$ the holes prefer to be on n.n. rungs so
that energy gain through the delocalization of a hole along a rung
is possible. The novel phenomena exhibited by undoped and doped spin
ladders have motivated a large number of theoretical studies. Several
ladder compounds have been discovered/synthesized to date. Detailed
information on the theoretical and experimental investigations may
be obtained from two exhaustive reviews on ladders$^{19,26}.$

\noindent \textbf{\large Haldane gap antiferromagnets}{\large \par}

We have so far been discussing SG systems with half-odd integer spins.
From the Lieb-Schultz-Mattis (LSM)$^{31}$theorem one can show that
the half-odd integer spin HAFM chain has a gapless excitation spectrum
in the infinite chain length limit. The theorem does not extend to
integer spin chains. Haldane$^{32},$ based on his analysis of the
nonlinear $\sigma$ model mapping of the large $S$ HAFM Hamiltonian
in 1d, conjectured that the HAFM spin chains with integer spins have
a gap in the excitation spectrum, i.e., are SG AFMs. The ground state
of an integer spin chain is disordered and the spin-spin correlation
function has an exponential decay. Haldane's conjecture has now been
verified both theoretically and experimentally$^{33}.$ The spin-$1$
HAFM Hamiltonian has the same ground state features as the spin-$1$
Affleck, Kennedy, Lieb, Tasaki (AKLT) Hamiltonian for which the so-called
valence bond solid (VBS) state is the exact ground state. Consider
a $1d$ lattice each site of which is occupied by a spin-$1.$ A spin-$1$
can be considered to be a symmetric combination of two spin-$\frac{1}{2}$'s.
In the VBS state, each spin-$\frac{1}{2}$ component of a spin-$1$
forms a singlet (VB) with a spin-$\frac{1}{2}$ at a neighbouring
site. The AKLT Hamiltonian is a sum over projection operators onto
spin $2$ for successive pairs of spins , i.e., 

\begin{equation}
H_{AKLT}=\sum_{i}P_{2}\left(\mathbf{S}_{i}+\mathbf{S}_{i+1}\right)\label{10}\end{equation}
\noindent The presence of a VB between each neighbouring pair of
sites implies that the total spin of each pair of spins cannot be
$2.$ Thus $H_{AKLT}$ acting on $\Psi_{VBS}$, the wave function
of the VBS state, gives zero. The eigenvalues of the projection operator
being positive, $\Psi_{VBS}$ is the ground state of $H_{AKLT}$ with
eigenvalue zero. The VBS state is spin-disordered with an exponentially
decaying spin-spin correlation function. One can, however, define
a non-local string order parameter which has a non-zero expectation
value in the VBS state$^{34}.$ The explicit form of the AKLT Hamiltonian
is given by

\begin{equation}
H_{AKLT}=\sum_{i}\left[\frac{1}{2}\left(\mathbf{S}_{i}.\mathbf{S}_{i+1}\right)+\frac{1}{6}\left(\mathbf{S}_{i}.\mathbf{S}_{i+1}\right)^{2}+\frac{1}{3}\right]\label{11}\end{equation}
\noindent The excitation spectrum of $H_{AKLT}$ cannot be determined
exactly. Variational calculations show the existence of a gap in the
excitation spectrum. Several Haldane gap (HG) AFMs have been discovered
so far. these include the $S=1$ compound $CsNiCl_{3}$$^{33},$ $Ni(C_{2}H_{8}N_{2})_{2}NO_{2}(ClO_{4})(NENP)$$^{33},$
$Y_{2}BaNiO_{5}$$^{35}$ and the $S=2$ compound ($2,2^{\prime}-$bipyridine)
trichloromanganese (III), $MnCl_{3}$ (bipyridine)$^{36}.$ Experiments
carried out on these compounds show that the VBS state provides the
correct physical picture for the true ground state$^{33}.$

The doped spin-$1$ compound $Y_{2-x}Ca_{x}BaNiO_{5}$ provides an
example of a QSL in $1d^{35}.$ The parent compound $Y_{2}BaNiO_{5}$is
a charge transfer insulator containing $Ni^{2+}(S=1)$ chains. The
ground state of the system is spin-disordered and the spin excitation
spectrum is separated by the HG from the ground state. The compound
is doped with holes on replacing the off-chain $Y^{3+}$ ions by $Ca^{2+}$ions.
The holes mostly appear in oxygen orbitals along the $NiO$ chains.
There is no evidence of metal-insulator transition but the dc-resistivity
$\rho_{dc}$ falls by several orders of magnitude. This indicates
that the holes are not fully mobile but delocalized over several lattice
spacings. Inelastic neutron scattering (INS) experiments reveal the
existence of new states within the HG. Several studies have been carried
out so far to explain the origin of the sub-gap states$^{37}.$ A
recent neutron scattering experiment$^{38}$ provides evidence for
an incommensurate double-peaked structure factor $S(q)$ for the sub-gap
states. The INS intensity is proportional to the structure factor.
For the pure compound, the structure factor $S(q),$ near the gap
energy of $9$ mev, has a single peak at the wave vector $q=\pi$
indicative of AFM correlations. For the doped compound, $S(q)$ has
an incommensurate double-peaked structure factor, for energy transfer
$\omega\sim3-7$ mev, with the peaks located at $q=\pi\pm\delta q.$
The shift $\delta q$ is found to have a very weak dependence on the
impurity concentration $x$ for $x$ in the range $x\epsilon\left[0.04,0.14\right].$
Evidence of incommensurate peaks has also been obtained in the underdoped
metallic cuprates. The peaks are four in number and occur at $(\pi\pm\delta q,\pi)$
and $(\pi,\pi\pm\delta q).$ The crucial difference from the nickelate
compound is that $\delta q$ is proportional to the dopant concentration
x. The incommensurability has been ascribed to inhomogeneous spin
and charge ordering in the form of stripes$^{23}.$ Malvezzi and Dagotto$^{39}$
have provided an explanation for the origin of spin incommensurability
in the hole-doped $S=1$ nickelate compound. They have shown that
a mobile hole generates AFM correlations between the spins located
on both sides of the hole and this is responsible for the spin incommensurability
seen in experiments. Xu et al.$^{38}$ have given a different explanation
for the origin of incommensurability. The holes doped into the QSL
ground state of the $S=1$ chain are located on the oxygen orbitals
and carry spin. They induce an effective ferromagnetic interaction
between the $Ni$ spins on both sides. The incommensurate peaks arise
because of the spin density modulations developed around the holes
with the size of the droplets controlled by the correlation length
of the QSL. Bose and Chattopadhyay$^{40}$ have developed a microscopic
theory of the origin of spin incommensurability in keeping with the
suggestions of Xu et al.

\noindent \textbf{\Large Quantum Phase Transitions}{\Large \par}

A quantum phase transition (QPT) occurs at $T=0$ and brings about
a qualitative change in the ground state of an interacting many body
system at a specific value $g_{c}$ of some tuning parameter $g.$
Examples of tuning variables include magnetic field $h,$ pressure
and dopant concentration. The origin of QPTs lies in quantum fluctuations
just as thermal fluctuations drive thermodynamic phase transitions.
In the case of second order thermodynamic phase transitions, the critical
point is characterised by scale invariance and a divergent correlation
length. Free energy and the different thermodynamic functions become
non-analytic at the critical temperature $T=T_{c}.$ The quantum critical
point is also associated with scale invariance and a divergent correlation
length with quantum fluctuations substituting for thermal fluctuations.
The ground state energy becomes non-analytic at the critical value
$g_{c}$ of the tuning parameter. If one of the phases is gapped,
the gap goes to zero in a power-law fashion as $g\rightarrow g_{c}.$
Quantum and thermal fluctuations are equally important in the so-called
quantum critical regime extending into the finite-$T$ part of the
$T$ versus $g$ phase diagram. The macroscopic physical properties
in this regime are in many cases independent of microscopic details.
A large number of theoretical and experimental studies has been carried
out on QPTs in condensed matter systems$^{41}.$ Here we focus on
a few specific examples of QPTs in SG AFMs. Organic spin ladder compounds
provide ideal testing grounds for theories of QPTs. Consider the phase
diagram of the AFM two-chain spin ladder in the presence of a magnetic
field $h.$ At $T=0$ and for $0<h<h_{c_{1}},$ the ladder is in the
SG phase. At $h=h_{c_{1}},$ there is a transition to the gapless
Luttinger Liquid phase with $g\mu_{B}h_{c_{1}}=\Delta,$ where $g$
is the Land\'{e} splitting factor, $\mu_{B}$ the Bohr magneton and
$\Delta,$ the magnitude of the SG$^{42}.$ At an upper critical field
$h_{c_{2}},$ there is another transition to the fully polarised ferromagnetic
state. Both $h_{c_{1}}$and $h_{c_{2}}$ are quantum critical points.
The compound $(C_{5}H_{12}N)_{2}CuBr_{4}$ has been identified as
a $S=\frac{1}{2}$ two-chain spin ladder in the strong coupling limit
$J_{R}=13.3K$ and $J=3.8K$ with $h_{c_{1}}=6.6T$ and $h_{c_{2}}=14.6T^{43}.$
The magnetization data exhibit universal scaling behaviour in the
vicinity of $h_{c_{1}}$ and $h_{c_{2}},$ consistent with theoretical
predictions. In the gapless regime $h_{c_{1}}<h<h_{c_{2}},$ the ladder
model can be mapped onto an XXZ chain, the thermodynamic properties
of which can be calculated exactly by the BA. The theoretically computed
magnetization $M$ versus magnetic field $h$ curve is in excellent
agreement with the experimental data. Other organic ladder compounds
exhibiting QPTs are $(5IAP)_{2}CuBr_{4}.2H_{2}O$$^{44}$ and $Cu_{2}(C_{5}H_{12}N_{2})_{2}Cl_{4}^{45}.$
For inorganic spin ladder systems, the value of $h_{c_{1}}$ is too
high to be experimentally accessible. The synthesis of organic ladder
compounds has paved the way for experimental observation of QPTs.

Another notable example of experimentally observed QPTs is that of
field-induced $3d$ magnetic ordering in low-dimensional SG AFMs,
$KCuCl_{3}$$^{3}$ and $TlCuCl_{3}$$^{4}$ which, as described earlier,
consist of networks of dimers. The dimer ground state is a singlet
for $h=0.$ The triplet $(S=1)$ excitation is separated from the
singlet ground state by an energy gap, $\Delta.$ Application of an
external magnetic field $h$ leads to Zeeman splitting of the triplet
excitation into three components: $S_{z}=+1,0,-1.$ At a critical
external magnetic field $h_{c_{1}},$ the lowest triplet component
becomes energetically degenerate with the ground state. This results
in a QPT at $h=h_{c_{1}}$to a $3d$ magnetically ordered state. The
critical point $h=h_{c_{1}}$ separates a gapped QSL state $(h<h_{c_{1}})$
from a field-induced magnetically ordered state $(h>h_{c_{1}}).$
The triplet components can be regarded as diluted bosons and the QPT
at $h=h_{c_{1}}$ can be treated as a Bose-Einstein condensation (BEC)
of low-lying magnons$^{46}.$ Support for this idea comes from experimental
findings on $TlCuCl_{3}.$ Recently, the excitation spectrum in the
magnetically ordered state of $TlCuCl_{3}$ has been determined using
INS$^{47}.$ The observed data are consistent with the theoretical
prediction of a gapless Goldstone mode characteristic of the BEC. 

In the BE condensed state, the state of each dimer is found to be
a coherent superposition of the singlet and the $S_{z}=+1$ triplet
states. The phase in the superposition specifies the orientation of
the staggered magnetization in the plane transverse to the magnetic
field direction. The number of magnons in the condensed state is not,
however, infinite as magnons cannot occupy the same sites in a spin
system due to a hard-core repulsion between them. The interaction
restricts the number of magnons to be large but finite. Recently,
there is a resurgence of research interest in BEC because of its experimental
realization in ultracold gases of dilute atoms. SG AFMs offer another
testing ground for theories related to BEC. The definitive evidence
of BEC in the compound $TlCuCl_{3}$ motivates the search for condensation
phenomena in other SG systems.

\noindent \textbf{\large Magnetization plateaus}{\large \par}

The magnetization curve of a low-$d$ AFM does not always show a smooth
increase in magnetization, from zero value to saturation, as the magnetic
field is increased in magnitude. In certain systems, the curve exhibits
plateaus at certain rational values of the magnetization per site
$m.$ The phenomenon is analogous to the quantum Hall effect in which
electrical resistivity exhibits plateaus as a function of the external
magnetic field. Oshikawa, Yamanaka and Affleck (OYA) derived a condition
for the occurrence of magnetization plateaus in quasi-$1d$ AFM systems
by generalising the LSM theorem to include an external magnetic field$^{48}.$
For general spin systems, the quantization condition can be written
as

\begin{equation}
S_{U}-m_{U}=integer\label{12}\end{equation}
\noindent where $S_{U}=nS,$ $n$ being the number of spins of magnitude
$S$ in unit period of the ground state and $m_{U}=nm$ is the magnetization
associated with the unit cell. The quantization condition is necessary
but not sufficient as not all plateaus predicted by the condition
exist in general. Hida$^{49}$ first predicted the existence of a
magnetization plateau at $m=\frac{1}{6}$ in the magnetization curve
of a $S=\frac{1}{2}$ AFM chain with period $3$ exchange coupling.
One can readily check that the quantization condition $(12)$ is obeyed
as in this case $S_{U}=\frac{3}{2}$ and $m_{U}=\frac{1}{2}$ $(n=3,S=\frac{1}{2},m=\frac{1}{6)}.$
A gapped phase is essential for the appearance of a plateau (magnetization
unchanging) in the $m$ versus $h$ curve. More than one plateau can
occur if there is more than one gapped phase as $h$ is changed. 

High-field measurements reveal the existence of magnetization plateaus
in several AFM compounds. The $S=\frac{1}{2}$ material $NH_{4}CuCl_{3}$
exhibits plateaus at $M=\frac{1}{4}$ and $\frac{3}{4}$ $(M=\frac{m}{S})^{50}.$
The $S=\frac{1}{2}$ SG AFM $SrCu_{2}(BO_{3})_{2}$ is the first example
of a $2d$ spin system for which magnetization plateaus have been
observed experimentally. The plateaus are obtained for $M=\frac{1}{3},\frac{1}{4}$
and $\frac{1}{8}^{51}.$ Momoi and Totsuka$^{52}$ have suggested
that the appearance of plateaus in $SrCu_{2}(BO_{3})_{2}$ is due
to a transition from a superfluid to a Mott insulating state of magnetic
excitations. As pointed out earlier, the triplet excitations in the
SS model, which describes $SrCu_{2}(BO_{3})_{2},$ are almost localized.
In the presence of a magnetic field and at special values of the magnetization,
the triplet excitations localize into a superlattice structure to
minimize energy so that the magnetization remains constant. As in
the cases of $KCuCl_{3}$ and $TlCuCl_{3},$ the compound $SrCu_{2}(BO_{3})_{2}$
is a network of coupled dimers. A triplet excitation created on a
dimer can propagate to a neighbouring dimer due to the inter-dimer
exchange interaction. The delocalization of triplets is similar to
that of electrons in crystals. In the presence of a magnetic field,
the triplet excitation is split into three components with the $S_{z}=+1$
component being the lowest in energy. The $S_{z}=+1$ excitations
can be regarded as bosons with a hard core repulsion. The repulsive
interaction arises from the $z$-component of the exchange interaction
and disallows the occupation of a single dimer by more than one boson.
The $xy$-part of the exchange interaction is responsible for the
hopping of the triplet excitation to neighbouring dimers. One thus
has a system of interacting bosons in which itinerancy competes with
localization. The transition from itinerancy to localization is analogous
to the Mott metal-insulator transition in electronic systems. If repulsive
interactions dominate, the triplet excitations (bosons) localize to
form a superlattice. A direct measurement of the magnetic superlattice
in $SrCu_{2}(BO_{3})_{2}$ has been made by Kodama et al.$^{52}$
using a high-field NMR facility. The superlattice corresponds to $M=\frac{1}{8}$
which requires a high magnetic field strength of $27$ $T$ for its
observation. Superlattice structures for higher $M$ values have not
been detected as yet because of the requirement of very high magnetic
fields. 

\noindent \textbf{\large Summary and Future Outlook}{\large \par}

SG AFMs exhibit a variety of novel phenomena the observation and interpretation
of which have been possible due to the intimate links between theory
and experiments. Rigorous theories and exactly solvable models have
predicted phenomena which were later verified experimentally. Quantum
magnetism is one of the few reseach areas in which rigorous theories
go hand in hand with experimental initiatives. We have illustrated
this interdependence through a few chosen examples. The MG model illustrates
the origin of SG due to the presence of further-neighbour interactions.
The SS and $J_{1}-J_{2}-J_{3}-J_{4}-J_{5}$ models extend the $1d$
model to $2d.$ Experimental realization of the SS model became a
reality about twenty years after the original theoretical proposal
was made. The AKLT model provides the correct physical picture of
the grounds states of integer spin chains. Knowledge of the ground
state has given rise to testable predictions which were later verified
experimentally. Spin ladders were originally studied as toy models
of strongly correlated systems. Theories of undoped and doped ladder
models motivated the search for real materials which led to success
in several instances. The OYA theorem provides the condition for the
existence of magnetization plateaus which is in agreement with experimental
results. SG AFMs further exhibit novel quantum phenomena which illustrate
how quantum effects influence ground and excited state properties.
We have discussed only a few of these in the present review. Two SG
AFM systems we have not described include the compounds $CaV_{4}o_{9}$$^{54}$
and the tellurate materials $Cu_{2}Te_{2}O_{5}X_{5}$$(X=Cl$ or $Br)^{55}.$The
first compound is defined on a $\frac{1}{5}$-depleted square lattice
which is a network of four-spin plaquettes connected by single links
(dimer bonds). Each square plaquette is in a resonating valence bond
(RVB) spin configuration in the ground state. In this case, the RVB
state is a linear superposition of two VB configurations. In one configuration,
the VBs (singlets) occupy the horizontal links of the square plaquette.
In the other configuration, the VBs occupy the vertical links. More
generally, a RVB state is a coherent linear superposition of VB states
and is a well-known example of a QSL. The tellurate materials can
be described as networks of spin tetrahedra. In both $CaV_{4}O_{9}$
and the tellurate materials, the existence of singlet excitations
in the spin gap has been reported as in the case of the kagom\'{e}
lattice HAFM. 

The QPTs considered in this review are brought about by tuning magnetic
field strengths. QPTs can also occur by the tuning of exchange interaction
strengths. In the case of the frustrated two-chain spin ladder model
described earlier, a first order QPT takes place at $\frac{J_{R}}{J}=\left(\frac{J_{R}}{J}\right)_{c}\simeq1.401$
from the rung dimer state to the Haldane phase of the $S=1$ chain$^{22}.$
Kolezhuk and Mikeska$^{56}$ have constructed generalised $S=\frac{1}{2}$
two-chain ladder models with two-spin and four-spin exchange couplings
for which the ground state can be determined exactly. QPTs to the
various phases are obtained by varying the exchange interaction strengths.
A lattice of coupled two-chain ladders provides another example of
a QPT which is brought about by tuning the inter-ladder exchange interaction
strength $\lambda^{41}.$ For $\lambda<\lambda_{c},$ the spin lattice
is in the SG phase. At $\lambda=\lambda_{c},$ a QPT to a long range
magnetically ordered state occurs. Most of the spin models considered
in this review exhibit QPTs at specific values of the exchange interaction
strengths. One might think that experimental observation of such QPTs
is not possible as exchange interaction strengths cannot be changed
at will. There is now reason to believe that exchange interaction
strengths can be controlled. Recent observations$^{57}$ of the superfluid
to Mott insulator transition in a system of ultracold atoms defined
in an optical lattice open up the exciting possibility of investigating
phenomena associated with interacting many body systems in a controllable
environment. The optical lattice is generated as a light-wave interference
pattern using several criss-crossing laser beams. The lattice is equivalent
to an energy landscape of mountains and valleys which can provide
the confining potential to trap individual atoms in separate valleys.
Proposals for the implementation of spin Hamiltonians in optical lattices
have been put forward with the aim to study the exotic quantum phases
of interacting spin systems$^{58}.$ The exchange interaction between
spins belonging to atoms in neighbouring valleys can be modified by
controlling the intensity, frequency and polarization of the trapping
light. Spin Hamiltonians of interest can be engineered through the
designing of appropriate optical lattice geometries. Practical implementation
of some of these ideas may be possible in the not too distant future.
Spin systems have recently been suggested as candidates for the realization
of quantum computation and communication protocols$^{59}.$The spin
systems considered so far include some SG AFMs like the MG and HG
chains, the two-chain spin ladder etc.$^{60}.$ Again, one anticipates
intense research activity in the coming years on such problems. To
sum up, the richness and vitality of the subject of SG AFMs are evident
in the challenging problems of current research interest and also
in the opening up of new avenues of research.

\newpage

\section*{Reference}

\begin{enumerate}
\item Bethe, H., Z. Physik, 1931, 71, 205 (English translation in the book
`The many-body Problem: An Encyclopedia of Exactly Solved Models in
One Dimension' (ed. Mattis, D.C.), World Scientific, 1993
\item Hase, M., Tarasaki, I. and Uchinokura, K., Phys. Rev. Lett., 1993,
70, 3651
\item Shiramura, W., Takatsu, K., Tanaka, H., Kamishima, K., Takahashi,
M., Mitamura, H. and Goto, T., J. Phys. Soc. Jpn., 1997, 66, 1900
\item Takatsu, K., Shimamura, W. and Tanaka, H., J. Phys. Soc. Jpn., 1997,
66, 1611
\item Ramirez, A. P., Ann. Rev. Mater. Sci., 1994, 24, 453
\item Chalker, J. T., Holdsworth, P. C. W. and Shender, E. F., Phys. Rev.
Lett., 1992, 68, 855
\item Huse, D. A. and Elser, V., Phys. Rev. Lett., 1988, 60, 2531
\item Bernu, B., Lhuillier, C. and Pierre, L., Phys. Rev. Lett., 1992, 69,
2590
\item Harrison, A. , Annual Reports on the Progress of Chemistry, 1992,
87A, 211
\item Villain, J., Bidaux, R., Carton, J. P. and Conte, R., Journal de Physique,
1980, 41, 1263; Shender, E. F., Sov. Phys. JETP, 1982, 56, 178
\item Lecheminant, P., Bernu, B., Lhuillier, C., Pierre, L. and Sindzingre,
P., Phys. Rev. B, 1997, 56, 2521
\item Mila, F., Eur. J. Phys., 2000, 21, 499
\item Majumdar, C. K. and Ghosh, D. K., J. Math. Phys., 1969, 10, 1388
\item Shastry, B. S. and Sutherland, B., Phys. Rev. Lett., 1981, 47, 964
\item Bose, I. and Mitra, P., Phys. Rev. B, 1991, 44, 443
\item Bhaumik, U. and Bose, I., Phys. Rev. B, 1995, 52, 12484
\item Shastry, B. S. and Sutherland, B., Physica B, 1981, 108, 1069
\item Miyahara, S. and Ueda, K., Phys. Rev. Lett., 1999, 82, 3701
\item Dagotto, E. and Rice, T. M., Science, 1996, 271, 618
\item Rice, T. M., Gopalan, S. and Sigrist, M., Europhys. Lett., 1993, 23
445
\item Bose, I. and Gayen, S., Phys. Rev. B, 1993, 48, 10653
\item Xian, Y., Phys. Rev. B, 1995, 52, 12485
\item Ornstein, J. and Millis, A. J., Science, 2000, 288, 468
\item Dagotto, E., Riera, J. and Scalapino, D. J., Phys. Rev. B, 1992, 45,
5744
\item Uehara, M., Nagata, T., Akimitsu, J., Takahashi, H., Mori, N. and
Kinoshita, K., J. Phys. Soc. Jpn., 1996, 65, 2764
\item Dagotto, E., Rep. Prog. Phys., 1999, 62, 1525
\item Bose, I. and Gayen, S., J. Phys.: Condens. Matter, 1994, 6, L405
\item Bose, I. and Gayen, S., J. Phys.: Condens. Matter, 1999, 11, 6427
\item Troyer, M., Tsunetsugu, H. and Rice, T. M., Phys. Rev. B, 1996, 53,
251
\item Rice, T. M., Z. Phys. B, 1997, 103, 165
\item Lieb, E., Schultz, T. D. and Mattis, D. C., Ann. Phys., 1961, 16,
407
\item Haldane, F. D. M., Phys. Rev. Lett., 1983, 50, 1153; Phys. Lett. A,
1983, 93, 464
\item Affleck, I., J. Phys.: Condens. Matter, 1989, 1, 3047
\item Affleck, I., Kennedy, T., Lieb, E. H. and Tasaki, H., Phys. Rev. Lett.,
1987, 59, 799; Commun. Math. Phys., 1988, 115, 477
\item Di Tusa, J. F., Cheong, S-W., Park, J.-H., Aeppli, G., Broholm, C.
and Chen, C. T., Phys. Rev. Lett., 1994, 73, 1857
\item Granroth, G. E., Meisel, M. W., Chaparala, M., Jolicoeur, T., Ward,
B. H. and Talham, D. R., Phys. Rev. Lett., 1996, 77, 1616
\item Penc, K. and Shiba, H., Phys. Rev. B, 1995, 52 R715; Dagotto, E.,
Riera, J., Sandvik, A. and Moreo, A., Phys. Rev. Lett., 1996, 76,
1731
\item Xu, G. et al., Science, 2000, 289, 419
\item Malvezzi, A. L. and Dagotto, E., Phys. Rev. B, 2001, 63, 140409
\item Bose, I. and Chattopadhyay, E., Int. J. Mod. Phys. B, 2001, 15, 2535
\item Sondhi, S. L., Girvin, S. M., Carini, J. P. and Shahar, D., Rev. Mod.
Phys., 1997, 69, 315; Sachdev, S., Science, 2000, 288, 475; Sachdev,
S., Quantum Phase Transitions, Cambridge University Press, Cambridge,
1999
\item Chaboussant, G. et al., Eur. Phys. J. B, 1998, 6, 167
\item Watson, B. C. et al., Phys. Rev. Lett., 2001, 86, 5168
\item Landee, C. P., Turnbull, M. M., Galeriu, C., Giantsidis, J. and Woodward,
F. M., Phys. Rev. B, 2001, 63, 100402
\item Chaboussant, G., Crowell, P. A., L\'{e}vy, L. P., Piovesana, O.,
Madouri, A. and Mailly, D., Phys. Rev. B, 1997, 55, 3046
\item Rice, T. M., Science, 2002, 298, 760; Nikuni, T., Oshikawa, M., Oosawa,
A. and Tanaka, H., Phys. Rev. Lett., 2000, 84, 5868; Normand, B.,
Matsumoto, M., Nohadani, O., Wessell, S., Haas, S., Rice, T. M. and
Sigrist, M., J. Phys.: Condens. Matter, 2004, 16 , S867
\item R\"{u}egg, Ch. et al., Nature, 2003, 423, 62
\item Oshikawa, M., Yamanaka, M. and Affleck, I., Phys. Rev. Lett., 1997,
78, 1984
\item Hida, K., J. Phys. Soc. Jpn., 1994, 63, 2359
\item Shiramura, W. et al., J. Phys. Soc. Jpn., 1998, 67, 1548
\item Kageyama, H. et al., Phys. Rev. Lett., 1999, 82, 3168; Onizuka, K.,
Kageyama, H., Narumi, Y., Kindo, K., Ueda, Y. and Goto, T., J. Phys.
Soc. Jpn., 2000, 69, 1016
\item Momoi, T. and Totsuka, K., Phys. Rev. B, 2000, 61, 3231
\item Kodama, K. et al., Science, 2002, 298, 395
\item Katoh, N. and Imada, M., J. Phys. Soc. Jpn., 1995, 64, 4105; Troyer,
M., Kontani, H. and Ueda, K., Phys. Rev. Lett., 1996, 76, 3822; Bose,
I. and Ghosh, A., Phys. Rev. B, 1997, 56, 3154
\item Johnsson, M., T\"{o}rnroos, K. W., Mila, F. and Millet, P., Chem.
Mater., 2000, 12, 2853; Brenig, W. and Becker, K. W., Phys. Rev. B,
2001, 64, 214413; Totsuka, K. and Mikeska, H-J., Phys. Rev. B, 2002,
66, 054435
\item Kolezhuk, A. K. and Mikeska, H-J., Int. J. Mod. Phys. B, 1998, 12,
2325
\item Greiner, M., Mandel, O., Esslinger, T., H\"{a}nsch, T. W. and Bloch,
I., Nature, 2002, 415, 39
\item Duan, L.-M., Demler, E. and Lukin, M. D., Phys. Rev. Lett., 2003,
91, 090402; Garcia-Ripoll, J. J., Martin-Delgado, M. A. and Cirac,
J. I., cond-mat/0404566; Pachos, J. K. and Plenio, M. B., quant-ph/0401106
\item Loss, D. and Di Vincenzo, D. P., Phys. Rev. A, 1998, 57, 120; Lidar,
D. A. and Wu, L.-A., Phys. Rev. Lett., 2002, 88, 017905; Bose, S.,
Phys. Rev. Lett., 2003, 91, 207901; Subrahmanyam, V., quant-ph/0307135
\item Bose, I. and Chattopadhyay, E., Phys. Rev. A, 2002, 66, 062320; Verstraete,
F., Martin-Delgado, M. A. and Cirac, J.I., Phys. Rev. Lett., 2004,
92, 087201; Fan, H., Korepin, V. and Roychowdhury, V., quant-ph/0406067;
Li, Y., Shi, T., Song, Z. and Sun, C. P., quant-ph/0406159 \newpage
\end{enumerate}
Figure Captions

\begin{enumerate}
\item Five types of interaction in the $J_{1}-J_{2}-J_{3}-J_{4}-J_{5}$
model. The successive interactions are  n.n., diagonal, n.n.n., knight's-move-distance-away
and further-neighbour-diagonal.
\item Four columnar dimer states. The dotted line represents a valence bond,
i.e., a singlet spin configuration.
\item The Shastry-Sutherland model. The n.n. and diagonal exchange interaction
strengths are $J_{1}$and $J_{2}$ respectively.
\item A two-chain spin ladder.\end{enumerate}

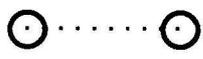
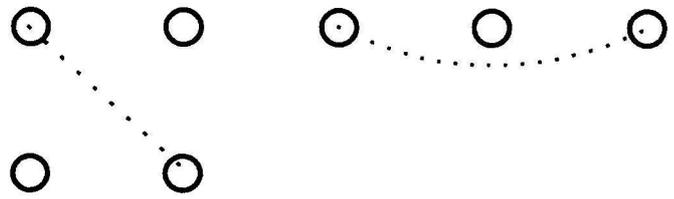
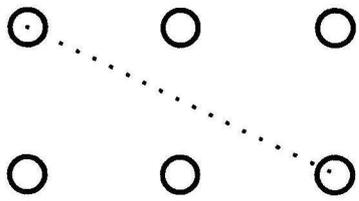
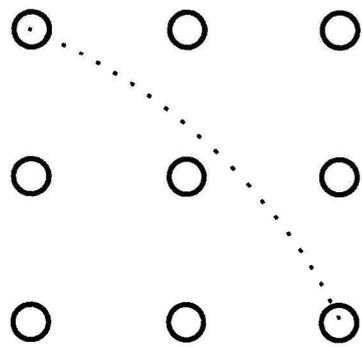
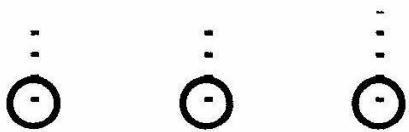
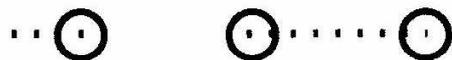
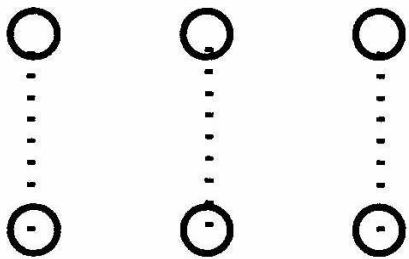
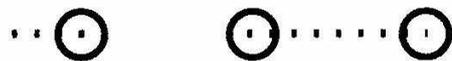
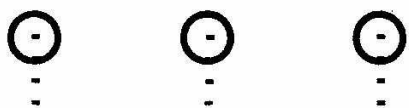
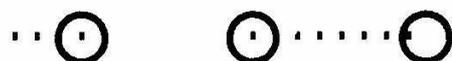
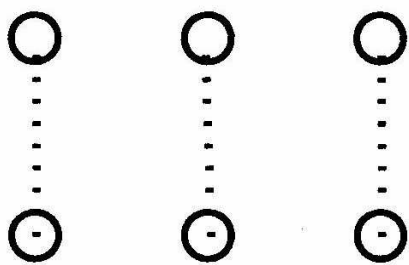
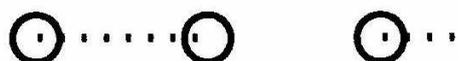
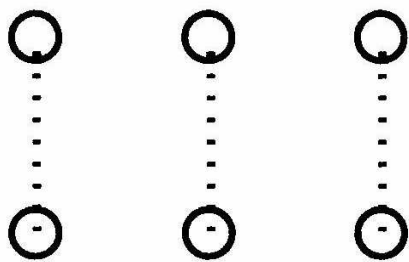
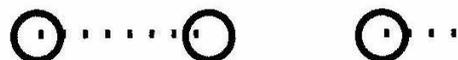
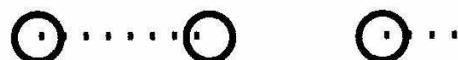
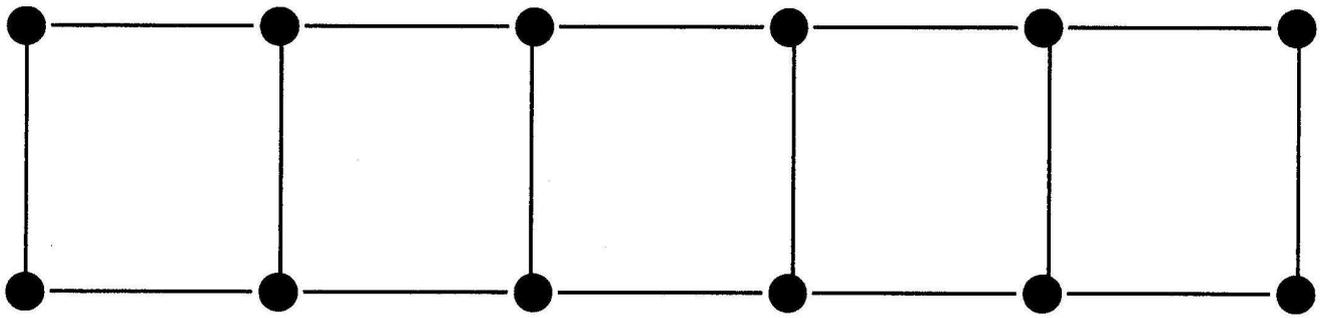
\end{document}